%% file: Jaewook_CPC2015.tex
\documentclass[final,5p,twocolumn,times]{elsarticle}

\biboptions{sort&compress}

\usepackage{amssymb}

\usepackage{braket}

\include{my_macro}

\usepackage[fleqn]{amsmath}
\usepackage{graphicx}
\usepackage{epstopdf}

\usepackage[english]{babel}

\newcommand{\Erf}{\text{Erf}}
\newcommand{\DT}{\Delta T}
\newcommand{\DTsub}{\DT_\text{sub}}
\newcommand{\Tmid}{T_\text{mid}}
\newcommand{\DL}{\Delta L}
\newcommand{\DLv}{\DL_\text{view}}
\newcommand{\lx}{\lambda_x}
\newcommand{\tlife}{\tau_\text{life}}
\newcommand{\WDTsub}{\mathcal{W}_{\DTsub}\lp t_i\rp}
\newcommand{\WDL}{\mathcal{W}_{\DL}}
\newcommand{\tac}{\tau_\text{ac}}
\newcommand{\normdist}{\mathcal{D}\lp\lx, v\rp}
\newcommand{\CaaSyn}{C_{a,a}^{\text{SYN}}}
\newcommand{\WDLmax}{\WDL^\text{max}}
\newcommand{\WDLmin}{{\WDL^\text{min}}}

\journal{Computer Physics Communications}

\begin{document}

\begin{frontmatter}



\title{Conditions for generating synthetic data to investigate characteristics of fluctuating quantities}


\author[a]{Jaewook~Kim}
\ead{ijwkim@kaist.ac.kr}
\author[b,c,d]{M.~F.~J.~Fox}
\author[b]{A.~R.~Field}
\author[e]{Y.U.~Nam}
\author[a]{Y.-c.~Ghim\corref{author}}
\ead{ycghim@kaist.ac.kr}
\cortext[author]{Corresponding author.}
\address[a]{\KAISTNQe}
\address[b]{\culham}
\address[c]{\oxford}
\address[d]{\merton}
\address[e]{\NFRI}

\begin{abstract}
Synthetic data describing coherent random fluctuations have widely been used to validate numerical simulations against experimental observations or to examine the reliability of extracting statistical properties of plasma turbulence via correlation functions. Estimating correlation time or lengths based on correlation functions implicitly assumes that the observed data are \textit{stationary} and \textit{homogeneous}. It is, therefore, important that numerically generated synthetic data also satisfy the stationary process and homogeneous state. Based on the synthetic data with randomly generated moving Gaussian shaped fluctuations both in time and space, the correlation function depending on the size of averaging time window is analytically derived. Then, the smallest possible spatial window size of synthetic data satisfying the stationary process and homogeneous state is proposed, thereby reducing the computation time to generate proper synthetic data and providing a constraint on the minimum size of simulation domains when using synthetic diagnostics to compare with experiment. This window size is also numerically confirmed with 1D synthetic data with various parameter scans.

\end{abstract}

\begin{keyword}
Synthetic data; Numerical domain; Turbulence; Statistical analysis



\end{keyword}

\end{frontmatter}



\setlength{\mathindent}{0pt}
\noindent

\section{Introduction}
\label{sec:intro} 
As the turbulence driven transport in a magnetically confined plasma exceeds the neoclassical transport level by at least an order of magnitude \cite{Carreras1997}, it is desirable to suppress the turbulence. For this purpose, we wish to understand the basic characteristics of the turbulence such as decorrelation rate and correlation lengths, and to perceive how they are correlated with equilibrium quantities, how they react back to these equilibrium quantities, and hopefully how they might be controlled \cite{Burrell_ExB,Diamond_zonalflows2005,Hahm_flowshear1995,Huld_coherentstructure1991,Mckee_velocimetry2004, ghim_prl_2013, ghim_nf_2014, Levinson1984}. Not being deterministic, turbulent structures must be studied based on the statistical grounds. Therefore, developing reliable statistical analyses to extract turbulence characteristics from the measured data is of paramount importance. For example, correlation functions can estimate correlation time and lengths of the turbulence, and the cross-correlation time delay method allows us to measure the velocity of pattern flows \cite{Ghim2012,Durst1992,Fonck1993}.

As numerical simulations and experimental diagnostics on plasma turbulence become more sophisticated, synthetic turbulence data generated from the simulations have been used to compare the results from simulations and experiments directly \cite{White_pop_2008, Shafer_PoP_2012, Field_ppcf_2014}. Turbulence synthetic data can also be used to examine the reliability of statistical techniques used to extract turbulence characteristics \cite{Bencze2005, Balazs2011, Ghim2012, Guszejnov2013}, i.e., turbulence characteristics extracted from the synthetic data using a statistical technique can be compared with the input parameters generating the synthetic data. 

The property of synthetic data themselves has not been thoroughly investigated so far. For instance,  as estimating correlation time and lengths using correlation functions from the measured data implicitly assumes that the data are stationary and homogeneous, synthetic data must also comply with the conditions of stationary process and homogeneous state. Stationary process means that low moments of fluctuating data such as mean and variance do not vary with time; while if they are unchanged in space, then the data are said to be homogeneous. To generate `true' stationary and homogeneous synthetic data, the simulation domain has to be infinitely large due to the finite correlation time and lengths of turbulent eddies. This is impractical. In practice, turbulent structures, or `eddies', are generated within a finite spatial domain and temporal domain. Therefore, for eddies which have a finite spatial and temporal extent, there are no sources from outside of these domains that contribute to the response within the domain (assuming that the boundary conditions are not periodic). Hence, these cause a spatial (and/or temporal) variation that leads to an inhomogeneous (non-stationary) correlation function.

\begin{figure}[t]
\includegraphics[width=\linewidth]{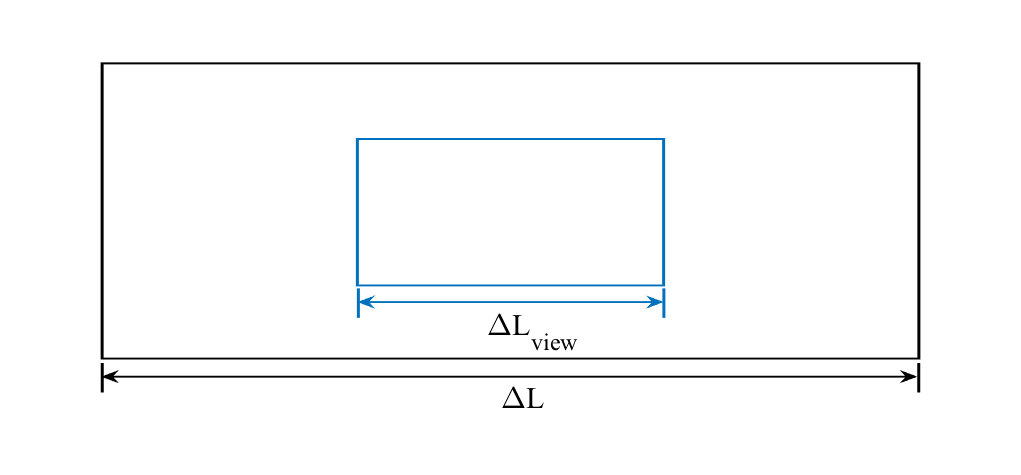}
\caption{A diagram depicting the total simulation domain $\DL$ and a smaller `viewing' domain (domain of interest) $\DLv$ where the generated synthetic data are stationary and homogeneous. Outside $\DLv$  the synthetic data may become non-stationary and/or non-homogeneous depending on how they are generated.}
\label{fig:window_size}
\end{figure}

In this paper, we thus provide the minimal size of required simulation domain $\DL$ given the `viewing' domain (domain of interest) $\DLv$ upon where one would apply statistical analyses as shown in \reffig{fig:window_size}. 
This means that generated synthetic data within $\DLv$ must be stationary and homogeneous, otherwise statistically calculated correlation functions may give us incorrect results. Of course, we wish to find the minimal $\DL$ so that we do not waste our computation resource. Or, for the case of local gyro-kinetic (GK) simulations where simulation domains $\DL$ are set, we provide the maximum possible $\DLv$ where the synthetic data can be valid for direct comparisons with experimental observations.

We first describe the mathematical model of a fluctuating quantity, or `eddy', such as density, temperature or potential in \refsec{sec:corr_fcn_eddy} and analytically derive correlation functions assuming that eddies are uniformly distributed in an infinitely large domain. In \refsec{sec:syn_data}, we provide the condition on the total simulation domain $\DL$ as a function of the `viewing' domain $\DLv$ and the size of the turbulent eddies, based on the derived correlation function such that the generated synthetic data satisfy stationarity and homogeneity. This condition is verified numerically using the 1D (in space) fluctuating synthetic data with various parameter scans. Note that even though we use 1D synthetic data, our arguments can be generalized to 3D as long as the basis vectors are orthogonal to each other. Our conclusion follows in \refsec{sec:conclusion}.

\section{Correlation function of `eddies'}
\label{sec:corr_fcn_eddy}
\subsection{Mathematical model of `eddies'}
\label{subsec:model_eddy}
In this section, we introduce a mathematical model describing real fluctuations as an ensemble of `eddies' -- its definition will follow soon -- based on which we derive the correlation function and generate synthetic data \cite{Bencze2005, Ghim2012, Balazs2011}. For simplicity we model the fluctuations in a 1D spatial domain. We represent our data at the spatial location $x=x_a$ as a function of time as
\begin{equation}
\label{eq:eddy_sum}
S_{a}(t)={\displaystyle \sum\limits _{i=1}^{N}S_{a_{i}}(t)},
\end{equation}
where $S_{a_i}\lp t\rp$ is the $i^\text{th}$ `eddy', and $N$ is the total number of eddies generated in the synthetic data.

\begin{figure}[t]
\includegraphics[width=\linewidth]{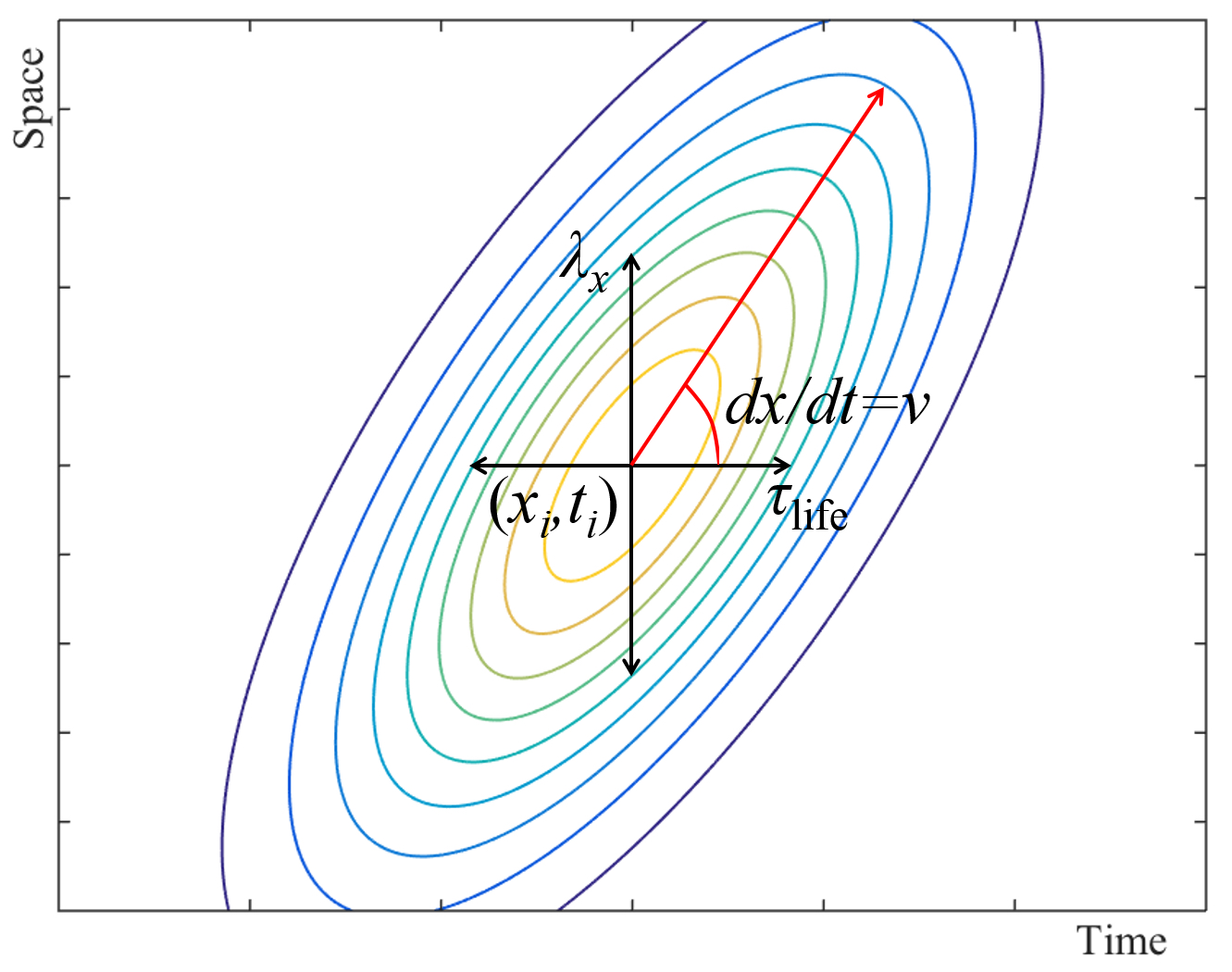}
\caption{An example of the contour of a single eddy in the space (ordinate) and time (abscissa) coordinate. The correlation length ($\lx$) and time ($\tlife$) in \refeq{eq:one_eddy} are also depicted. The slope of the red line is the velocity of the eddy.}
\label{fig:Contour-of-single}
\end{figure}

We have many different possibilities on what mathematical form $S_{a_i}\lp t\rp$ would take. Inspired by the experimental observations on ion-scale density fluctuations \cite{Mckee2003,Fonck1993}, we model that eddies are Gaussian shaped in both time and space:
\begin{equation}
\label{eq:one_eddy}
S_{a_i}\lp t\rp = A_i\exp\lsb -\dfrac{\lp t-t_i\rp^2}{2\tlife^2}-\dfrac{\lp x_a-v\lp t-t_i\rp - x_i \rp^2}{2\lx^2} \rsb.
\end{equation} 
Coherent properties of each eddy in space and time are parameterized by the characteristic spatial scale ($\lx$) and the characteristic temporal scale ($\tlife$). The $i^\text{th}$ eddy has a maximum amplitude $A_i$ at $x=x_i$ and $t=t_i$. Further, we allow an eddy to move with the velocity of $v$. Note that our model eddy does not contain the wave-like structures \cite{Ghim2012}, and we justify it by arguing that we are primarily interested in the envelope of eddies. Here, $A_i$ is selected from a normal distribution with zero mean and variance of $A^2$; whereas $x_i$ and $t_i$ are randomly selected from uniform distributions:
\begin{equation}
\begin{aligned}
\label{eq:xtA_dist}
P(t_{i})=&\begin{cases}
\dfrac{1}{\Delta T} & \mbox{if }-\dfrac{\Delta T}{2}\leq t_{i}\leq\dfrac{\Delta T}{2}\\
\\
\hfil0 & \hfil\mbox{otherwise}
\end{cases} \\
& \\
P(x_{i})=&\begin{cases}
\dfrac{1}{\Delta L} & \mbox{if }-\dfrac{\Delta L}{2}\leq x_{i}\leq\dfrac{\Delta L}{2}\\
\\
\hfil0 & \hfil\mbox{otherwise}
\end{cases} \\
&\\
P(A_{i})=&\frac{1}{\sqrt{2\pi}A}\exp\lsb -\frac{A_i^2}{2A^2} \rsb,
\end{aligned}
\end{equation}
where $P\lp t_i\rp$, $P\lp x_i\rp$ and $P\lp A_i \rp$ are the probabilities of obtaining $t_i$, $x_i$ and $A_i$, respectively. $\DT$ and $\DL$ are the total simulation domains in time and space, respectively (as in \reffig{fig:window_size} for $\DL$). Furthermore, to make sure that eddies do not occur too frequently or too rarely, we define a spatio-temporal filling factor $F$ \cite{Ghim2012}. We determine the total number of eddies ($N$) generated in a set of synthetic data such that the following expression is satisfied:
\begin{equation}
\label{eq:filling_factor}
F=N\left({\displaystyle \frac{\lx}{\DL}}\right)\left({\displaystyle \frac{\tlife}{\DT}}\right)\sim\mathcal{O}\lp 1\rp.
\end{equation}
\reffig{fig:Contour-of-single} shows an example of the contour of a generated eddy in the spatial and temporal coordinates.

\subsection{Correlation function of stationary and homogeneous fluctuating data}
\label{subsec:deriv_corr_func}
As many kinds of statistical analyses are performed on the data based on the stationary and homogeneous assumptions, we let $\DT$ and $\DL$ to be infinite to make sure that our model data are stationary and homogeneous. To analytically calculate the correlation function following Tal et al. \cite{Balazs2011} between two spatial positions, $x_a$ and $x_b$, as a function of time delay $\tau$, we average the signals $S_a\lp t\rp$ and $S_b\lp t\rp$ over the `subtime' window $\DTsub$:
\begin{equation}
\label{eq:corr_ab_definition}
\begin{aligned}
C_{a,b}(\tau)= & \overline{(S_{a}(t)-\overline{S_{a}})(S_{b}(t+\tau)-\overline{S_{b}})}\\
\approx & \overline{S_{a}(t)S_{b}(t+\tau)}-\overline{S_{a}}\,\overline{S_{b}},
\end{aligned}
\end{equation}
where the approximation is allowed because the data are stationary \cite{Balazs2011,randomdata}. The overline means the time average of the signal over the subtime window $\DTsub$:
\begin{equation}
\begin{aligned}
\label{eq:overline_def}
\overline{S_{a_{i}}(t)S_{b_{j}}(t+\tau)} = & \dfrac{1}{\DTsub}{\displaystyle \int_{-\frac{\DTsub}{2}+\Tmid}^{\frac{\DTsub}{2}+\Tmid}S_{a_{i}}(t)S_{b_{j}}(t+\tau)dt} \\
= & C_{a_{i},b_{j}}(\tau), 
\end{aligned}
\end{equation}
where $T_\text{mid}$ is the time at the middle of the selected subtime window. Note that $C_{a_{i},b_{j}}(\tau)-\overline{S}_{a_i}\,\overline{S}_{b_j}$ is the correlation function between the $i^\text{th}$ eddy at $x=x_a$ and the $j^\text{th}$ eddy at $x=x_b$. 

As a set of total fluctuation data is the sum of all eddies as in \refeq{eq:eddy_sum}, we can expand $C_{a, b}$ as
\begin{equation}
\label{eq:corr_ab_expand}
C_{a,b}=  \sum_{i}C_{a_{i},b_{i}}+\sum_{i}\sum_{j\neq i}C_{a_{i},b_{j}} -\sum_{i}\overline{S}_{a_{i}}\,\overline{S}_{b_{i}}-\sum_{i}\sum_{j\neq i}\overline{S}_{a_{i}}\,\overline{S}_{b_{j}}.
\end{equation}
Finally, by averaging correlation functions estimated from many subtime windows, we get the ensemble averaged correlation function as
\begin{equation}
\label{eq:corr_ab_ensemble}
\begin{aligned}\braket{C_{a,b}}= & N\braket{C_{a_{i},b_{i}}}+N(N-1)\braket{C_{a_{i},b_{j}}}\\
 & -N\braket{\overline{S}_{a_{i}}\overline{S}_{b_{i}}}-N(N-1)\braket{\overline{S}_{a_{i}}}\braket{\overline{S}_{b_{j}}},
\end{aligned}
\end{equation}
where the second and the fourth terms on the right-hand-side cancel out. Furthermore, these two terms are independently zero if the mean of $A_i$ in \refeq{eq:one_eddy} is zero.

\refeq{eq:corr_aibi} shows $C_{a_i, b_i}$ at the time delay $\tau=0$ as one would do to attain the correlation length from the ensemble averaged correlation function.
\begin{equation}
\begin{aligned}
\label{eq:corr_aibi}
C_{a_{i},b_{i}}= & \dfrac{1}{\DTsub}{\displaystyle \int_{-\frac{\DTsub}{2}+\Tmid}^{\frac{\DTsub}{2}+\Tmid}S_{a_{i}}(t)S_{b_{i}}(t)dt}\\
= & A_i^2\dfrac{\sqrt\pi}{2}\dfrac{\tac}{\DTsub}\exp\lsb -\dfrac{\tac^2}{2\lx^2/v^2}\dfrac{\lp x_a-x_b\rp^2}{2\lx^2} \right. \\
&\qquad\qquad \left. -\dfrac{\tac^2}{\tlife^2}\dfrac{\lp x_a-x_i\rp^2+\lp x_b-x_i\rp^2}{2\lx^2}\rsb\WDTsub\\\\
\approx & \begin{cases}
A_i^2\sqrt\pi\dfrac{\tac}{\DTsub}\exp\lsb -\dfrac{\tac^2}{2\lx^2/v^2}\dfrac{\lp x_a-x_b\rp^2}{2\lx^2} \right. \\
\qquad\qquad\qquad\quad \left. -\dfrac{\tac^2}{\tlife^2}\dfrac{\lp x_a-x_i\rp^2+\lp x_b-x_i\rp^2}{2\lx^2}\rsb \\
\\
\quad\text{for } -\dfrac{\DTsub}{2}+\Tmid + \gamma_i   \leq t_{i}\leq\dfrac{\DTsub}{2}+\Tmid+\gamma_i
\\
\\
\qquad\qquad \qquad\ 0 \qquad\qquad \qquad\qquad \text{otherwise},
\end{cases}
\end{aligned}
\end{equation}
where $\tac$ is the usual auto-correlation time of eddies in the lab frame defined as \cite{Bencze2005}
\begin{equation}
\label{eq:tauc}
\tac = \dfrac{\lx\tlife}{\sqrt{\lx^2 + \tlife^2 v^2}},
\end{equation}
and $\gamma_i$ is
\begin{equation}
\label{eq:gamma}
\gamma_i = \dfrac{\tac^2}{\lx^2}v\lsb x_i - \dfrac{x_a+x_b}{2} \rsb.
\end{equation}
Here, $\WDTsub$ which acts like a weighting factor is a function containing the error function $\Erf\lp \rp$ defined as
\begin{equation}
\begin{aligned}
\label{eq:corr_erf}
\WDTsub=&\Erf\lsb\dfrac{1}{\tac}\left( \Tmid+\dfrac{\DTsub}{2}-t_{i}+\gamma_i\right) \rsb- \\
&\Erf\lsb\dfrac{1}{\tac}\left( \Tmid-\dfrac{\DTsub}{2}-t_{i}+\gamma_i\right) \rsb.\\
\end{aligned}
\end{equation}
The approximation in the last step in \refeq{eq:corr_aibi} is taken by assuming large $\DTsub$ such that $\WDTsub\approx 2$.

Once we have $C_{a_i, b_i}$, we calculate $\lab C_{a_i, b_i}\rab$ by taking an ensemble average with the probability density functions defined in \refeq{eq:xtA_dist}:
\begin{equation}
\label{eq:corr_ensemble}
\begin{aligned}
\left<C_{a_{i}b_{i}}\right>=& {\displaystyle \int_{-\infty}^{\infty}dA_i\displaystyle \int_{-\infty}^{\infty}dt_i{\displaystyle \int_{-\infty}^{\infty} dx_i P(x_{i})P(t_{i})P(A_i)C_{a_{i}b_{i}}}}\\
=&\lp A^2\dfrac{\pi}{2}\dfrac{\tlife}{\DT}\dfrac{\lx}{\DL}\rp\exp\left[-\dfrac{(x_{a}-x_{b})^{2}}{2\lp\sqrt 2\lx\rp^{2}}\right]\WDL\\
\approx&\lp A^2 \pi\dfrac{\tlife}{\DT}\dfrac{\lx}{\DL}\rp\exp\left[-\dfrac{(x_{a}-x_{b})^{2}}{2\lp\sqrt 2\lx\rp^{2}}\right],
\end{aligned}
\end{equation}
where $\WDL$ plays the similar role as $\WDTsub$ did for \refeq{eq:corr_aibi} containing the error function $\Erf\lp \rp$ defined as
\begin{equation}
\label{eq:ensemble_corr_erf}
\WDL=\Erf\lsb\dfrac{x_{a}+x_{b}+\DL}{2\sqrt{\lx^{2}+\tlife^{2}v^{2}}}\rsb-\Erf\lsb\dfrac{x_{a}+x_{b}-\DL}{2\sqrt{\lx^{2}+\tlife^{2}v^{2}}}\rsb.
\end{equation}
Similar to what we did for $\WDTsub$, we approximate $\WDL\approx 2$ in \refeq{eq:corr_ensemble} by assuming infinitely large $\DL$. Note that we also have large $\DT$ because $\DT\gg\DTsub$, and $\DTsub$ is assumed to be large from \refeq{eq:corr_aibi}. After another lengthy algebraic calculation, we find that $\lab \overline{S}_{a_{i}}\overline{S}_{b_{i}} \rab$ in \refeq{eq:corr_ab_ensemble} is 
\begin{equation}
\label{eq:sig_ensemble}
\lab \overline{S}_{a_{i}}\overline{S}_{b_{i}} \rab \approx \lp A^2 \pi\dfrac{\tlife}{\DT}\dfrac{\lx}{\DL}\rp\lp2\sqrt\pi\dfrac{\tac}{\Delta T_{sub}}\rp \exp\lsb-\dfrac{(x_{a}-x_{b})^{2}}{4(\lambda_{x}^{2}+\tau_{life}^{2}v^{2})}\rsb,
\end{equation}
by applying the same assumptions, i.e., large $\DL$, $\DT$ and $\DTsub$. 

Collecting \refeq{eq:corr_ensemble} and \refeq{eq:sig_ensemble}, we finally obtain the ensemble averaged correlation value or the expected correlation value between the signals at $x=x_a$ and $x_b$ at the time delay $\tau=0$:
\begin{equation}
\label{eq:corr_anal_final}
\begin{aligned}
\lab C_{a,b} \rab \approx& A^2\pi N\frac{\lx}{\DL}\frac{\tlife}{\DT}\lp\exp\lsb -\frac{\lp x_a -x_b\rp^2}{2\lp\sqrt 2\lx\rp^2}\rsb \right. \\
& \qquad\qquad \left. -2\sqrt\pi\frac{\tac}{\DTsub}\exp \lsb-\frac{\lp x_a-x_b\rp^2}{4\lp\lx^2+\tlife^2 v^2 \rp} \rsb\rp \\
\approx&A^2\pi \lp\exp\lsb -\frac{\lp x_a -x_b\rp^2}{2\lp\sqrt 2\lx\rp^2}\rsb \right. \\
& \qquad\qquad \left. -2\sqrt\pi\frac{\tac}{\DTsub}\exp \lsb-\frac{\lp x_a-x_b\rp^2}{4\lp\lx^2+\tlife^2 v^2 \rp} \rsb\rp,
\end{aligned}
\end{equation}
where we use \refeq{eq:filling_factor} to get the last line. This is our final form of the ensemble averaged correlation function at the time delay $\tau=0$ with the assumptions of infinitely large domains and is used for numerical comparisons in \refsec{sec:syn_data}.

The first term has a Gaussian form as the shape of an individual eddy is set to be Gaussian (see \refeq{eq:one_eddy}). Note that the shape of a correlation function can be determined via a convolution of an individual eddy function with itself. In general, a correlation length is estimated by fitting correlation values $\lab C_{a,b}\rab$ to a Gaussian function with the knowledge of $x_a-x_b$. Such a fitting procedure, thus, implicitly ignores the second term in \refeq{eq:corr_anal_final} originated from $\lab \overline{S}_{a_{i}}\overline{S}_{b_{i}} \rab$. Discussing the finite effect of the second term in estimating $\lx$ is not within the scope of this paper. However, we briefly mention that ignoring the second term can be justified for $\tlife^2/\DTsub^2\ll 1$ given $\lx^2 \gg \tlife^2 v^2$, or for $\lp \lx/v\rp^2/\DTsub^2\ll 1$ given $\lx^2 \ll \tlife^2 v^2$. For the case of $\lx^2 \sim \tlife^2 v^2$, the situation becomes a bit more complicated, but large enough $\DTsub$ allows one to ignore the second term effect as well. Furthermore, an astute reader may realize that estimating the correlation length by fitting a Gaussian function to the first term overestimates the characteristic spatial scale $\lx$ by a factor of $\sqrt 2$. Nonetheless, such an overestimation may not become problematic if one is interested in the `scaling' of the correlation lengths rather than absolute quantities. In fact, as we do not have the first principle argument on what form of the correlation function, i.e., exponential function, Gaussian function, power-law function, etc, should be fitted to the experimental turbulence data, speaking of an absolute correlation length from the fitting must be done with great care. Perhaps, it is worth to consider the Gaussian process \cite{06MIT_Carl} to fit the data since we do not have well defined a prior knowledge on a plasma turbulence model function.

\section{Synthetic data}
\label{sec:syn_data}
\subsection{Conditions for generating stationary and homogeneous synthetic data}
\label{subsec:cond_syn_data}
In \refsec{sec:corr_fcn_eddy}, we have derived the correlation function assuming that arguments inside the $\Erf\lp \rp$ are large enough so that $\Erf\lp \rp$ returns $\pm 1$ depending on the sign of arguments. It is important to realize that this assumption is not used for a mere simplification of equations, rather it is the consequence of data with finite coherent structures, i.e., $\tlife\ne0$ or $\lx\ne0$, being homogeneous and stationary. To be quantitative, we argue that $\abs{\Erf\lp x\rp}\ge 0.995$ (or $\abs x\ge 2$) is the condition satisfying the assumption we have made, i.e., $\abs{\Erf\lp x\rp}\rightarrow 1$.

\begin{figure}[t]
\includegraphics[width=\linewidth]{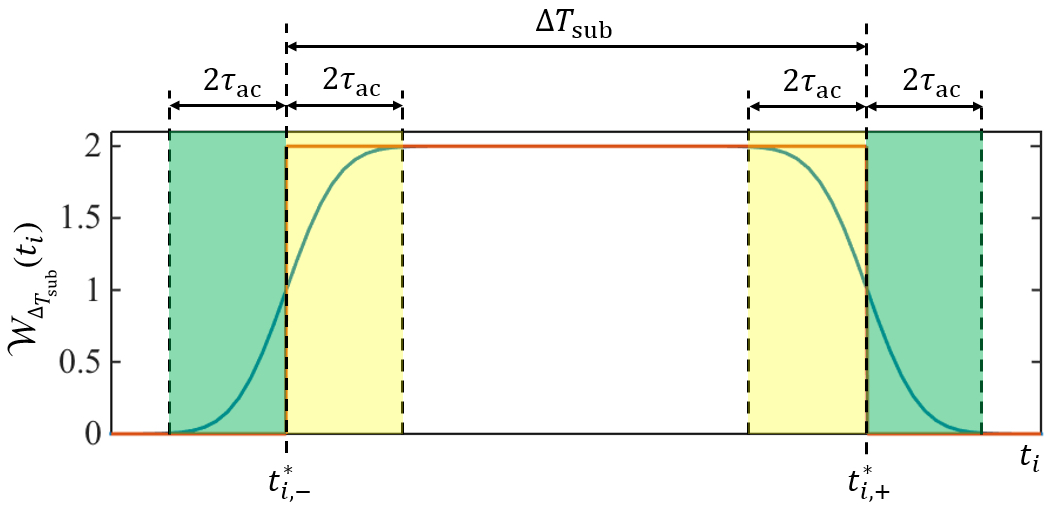}
\caption{An illustration of exact (blue) and approximated (red) $\WDTsub$. At $t_i=t_{i,\pm}^*$, we have $\WDTsub\approx 1$. In yellow (green) shaded regions whose widths are $2\tac$, the approximated $\WDTsub$ over(under)-estimates the exact value. We wish to minimize the fraction of the shaded regions in $\DTsub$.}
\label{fig:error_function_part_a}
\end{figure}

First, we find the condition on how large $\DTsub$ has to be by referring to \refeq{eq:corr_erf} as the $\DTsub$ appears inside the $\Erf\lp \rp$:
\begin{equation}
\label{eq:cond_DTsub}
\dfrac{\left| t_{i,\pm}^*-t_i\right|}{\tac}\ge 2,
\end{equation}
where $t_{i,\pm}^*$ is defined as 
\begin{equation}
\label{eq:tistar}
t_{i,\pm}^* = \Tmid \pm \frac{\DTsub}{2} + \gamma_i,
\end{equation}
and by definition we have $\DTsub = t_{i,+}^*-t_{i,-}^*$. Notice that at $t_i=t_{i,\pm}^*$, we have $\WDTsub\approx 1$ as shown in \reffig{fig:error_function_part_a} where the exact $\WDTsub$ (blue) and its approximation (red), i.e., $\WDTsub\approx 2$, are plotted. Considering the exact and approximated $C_{a_i, b_i}$ in \refeq{eq:corr_aibi}, we find that the value of approximated $C_{a_i, b_i}$ underestimates the true value in the green shaded region; while it overestimates in the yellow shaded region in \reffig{fig:error_function_part_a}. The width of each shaded region is approximately $2\tac$. Thus, to obtain the correct correlation values, it is necessary to have $\DTsub$ much larger than $\tac$ such that the fraction of under- or over-estimated regions are small. Casting this idea into a quantitative form, it states that
\begin{equation}
\label{eq:varepsilon}
\varepsilon\equiv\dfrac{2\tac}{\DTsub}\ll 1.
\end{equation}
It is well known that the size of averaging time window $\DTsub$ must be much larger than the auto-correlation time $\tac$ to obtain correct correlation functions, and here we have provided a quantitative rationale behind such a criterion. Note that this criterion does not guarantee the stationary process of the data.

\begin{figure}[t]
\includegraphics[width=\linewidth]{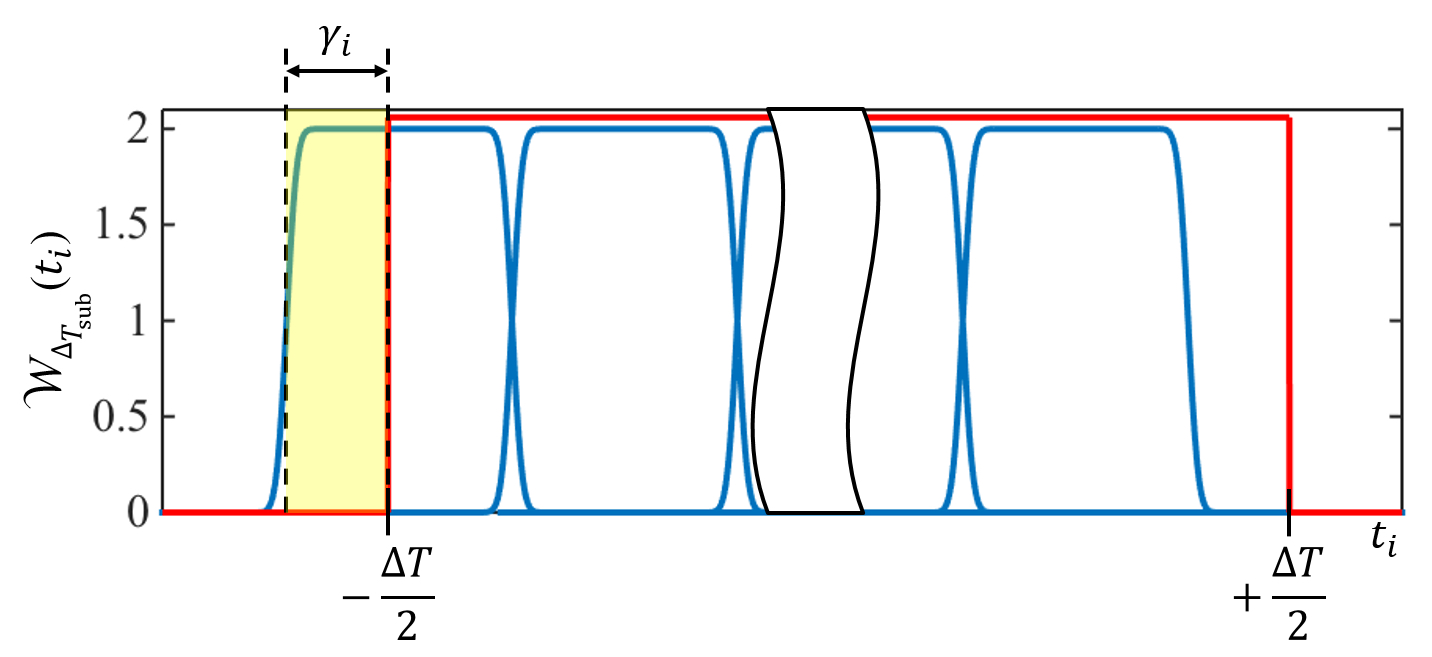}
\caption{Blue lines show the multiple $\WDTsub$'s with the widths of $\DTsub$ inside the total $\DT$ window set by the $P\lp t_i\rp$ depicted by the red line. All the $\WDTsub$'s are shifted by an amount of $\gamma_i$. The correlation value is underestimated for the far left subtime window containing the yellow shaded region. We wish to have the size of $\gamma_i$ negligible compared to $\DT$.}
\label{fig:error_function_part_b}
\end{figure}

It is obvious that the condition of $\DT\gg\DTsub$ must be well satisfied for the ensemble average defined in \refeq{eq:corr_ensemble}, and it is typically the case that the synthetic data are generated with a large enough $\DT$. However, for the sake of completeness, we provide more quantitative criterion on $\DT$ such that the true correlation value can be estimated from the synthetic data generated with the `finite' $\DT$. We realize that $\WDTsub$ is shifted by an amount of $\gamma_i$ as depicted in \reffig{fig:error_function_part_b} where the blue lines show the shifted $\WDTsub$, and the red line shows the uniform probability density function of $t_i$, $P\lp t_i\rp$. The $\WDTsub$ in the middle of $P\lp t_i\rp$ would not cause any problems on estimating correlation values, but the $\WDTsub$ at the edge of $P\lp t_i\rp$ may cause the underestimation of the true correlation value. This is because no eddies are generated in the shaded region, i.e., $P\lp t_i\rp$=0, while the size of the averaging subtime window is kept to be $\DTsub$ as other subtime windows. This violates the condition of data being stationary. To minimize this effect of underestimation we, thus, need to have the maximum of  $\abs{\gamma_i}$ much smaller than $\DT$. 

From the definition of $\gamma_i$ in \refeq{eq:gamma}, the maximum possible value of $\abs{x_i}$ is the $\DL/2$ from the probability density function of $x_i$, i.e., $P\lp x_i\rp$, and that of $\abs{x_a+x_b}$ is $\DLv$ since $x=x_a$ and $x_b$ are the `measurement' positions where we apply statistical analyses. Note that this is valid for the center of $\DLv$ coinciding with that of $\DL$. Thus, we obtain the condition on $\DT$, in addition to $\DTsub\ll\DT$, such that the synthetic data satisfy the stationary process:
\begin{equation}
\label{eq:gamma_constraints}
\begin{aligned}
\text{max}\lp\abs{\gamma_i}\rp=\dfrac{\tac^2}{\lx^2}v\lsb\dfrac{\DL}{2}+\dfrac{\DLv}{2}\rsb \ll&\DT \quad \text{ and} \\
\DTsub \ll&\DT,
\end{aligned}
\end{equation}
which depends on the spatial domain size due to the finite velocity of eddies.

To determine the minimal size of spatial domain $\DL$ given the `viewing' domain $\DLv$, we need to examine \refeq{eq:corr_ensemble} where the approximation of $\WDL\approx 2$ is supported by assuming large $\DL$. Again, this assumption is regarded to be well satisfied if the argument is larger than two:
\begin{equation}
\label{eq:DL_in_erf}
\left|\dfrac{x_{a}+x_{b}\pm\DL}{2\sqrt{\lx^{2}+\tlife^{2}v^{2}}}\right|\ge2.
\end{equation}
Since $\DLv\ge \left| x_{a}+x_{b}\right|$, we find a criterion on $\DL$ as
\begin{equation}
\label{eq:DL_condition}
4\sqrt{\lx^{2}+\tlife^{2}v^{2}}+\DLv \le \DL,
\end{equation}
and this condition guarantees that the ensemble averaged correlation values do not depend on the `measurement' positions, hence the synthetic data satisfy the homogeneous state.

If we delve into the structure of \refeq{eq:corr_ensemble} deeper, one may raise a question: what happens if $\WDL$ is almost constant at a value other than two within the `viewing' domain $\DLv$? If this happens, then we realize that $\lab C_{a_i,b_i}\rab$ does not depend on the spatial position as long as the distances between the two points, $\abs{x_a-x_b}$, are the same. This consequence may be argued for the data being homogeneous without satisfying the condition on $\DL$ set by \refeq{eq:DL_condition}. We provide more detailed explanation on this in \ref{app:homogeneous_state}.

\subsection{Simulation results with parameter scans}
\label{sec:sim_result_param_scan}
\begin{figure}[t]
\includegraphics[width=\linewidth]{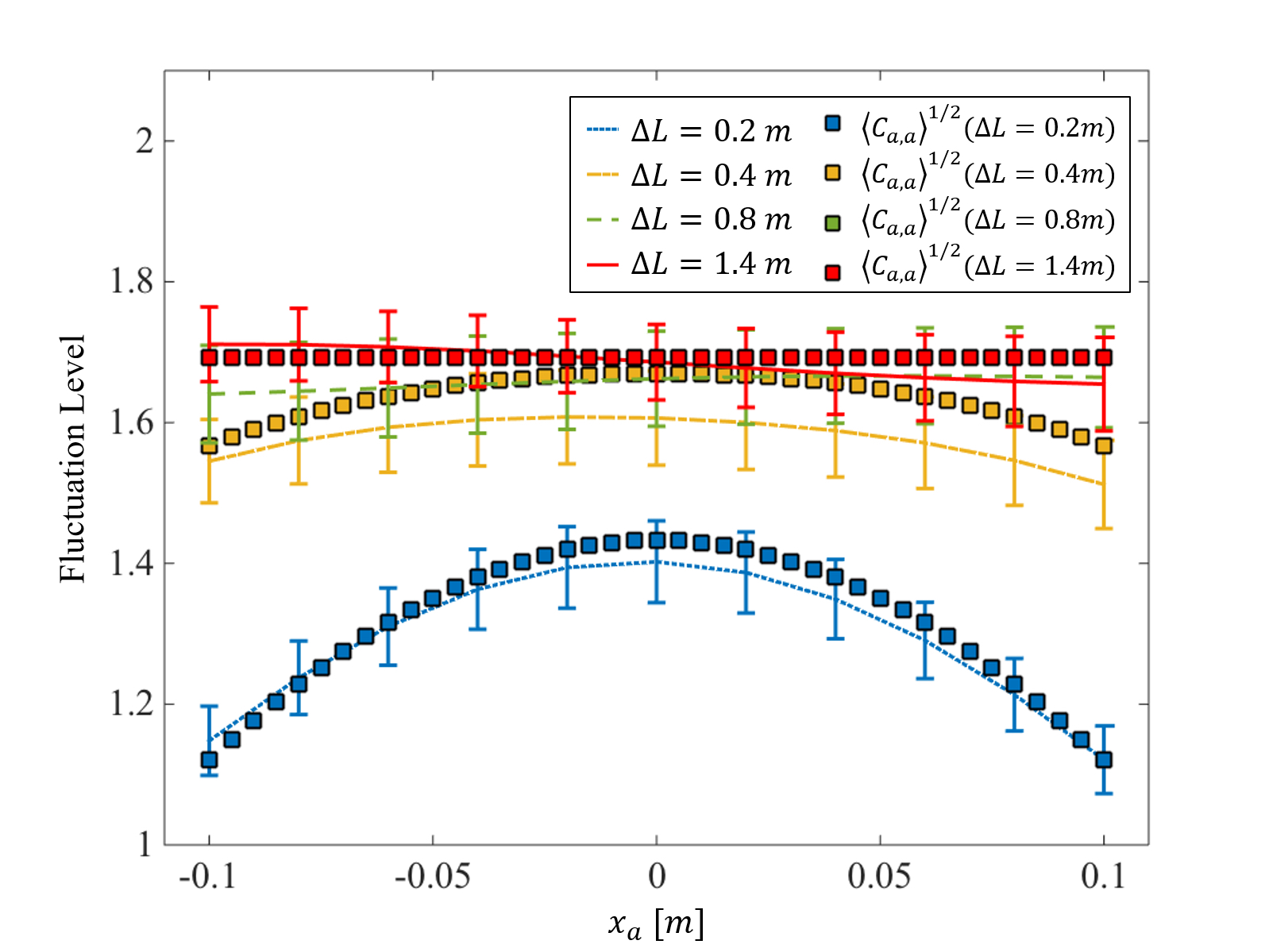}
\caption{Analytically calculated fluctuation levels of the synthetic data $\lab C_{a, a}\rab^{1/2}$ (squares) using \refeq{eq:corr_anal_final} with the actual values of $\WDL$, i.e., no approximation of $\WDL$ to two, and numerically estimated fluctuation levels $\lab\CaaSyn\rab^{1/2}$ from the synthetic data as a function of spatial position $x_a$ within $\DLv$ for $\DL=0.2$ (blue dot), $0.4$ (yellow dash dot), $0.8$ (green dash) and $1.4$ m (red line). Uncertainties represent the $95$\% confidence level of estimating fluctuation levels. Here, $\DL\ge 0.7$ m is the condition for the data to be homogeneous.}
\label{fig:Fluctuation-level}
\end{figure}

We now have the conditions on the sizes of temporal ($\DT$) and spatial ($\DL$) simulation domains given the viewing domain size ($\DLv$), correlation length, lifetime (decorrelation rate) and velocity of eddies such that the synthetic data are stationary and homogeneous. If the proposed conditions, \refeq{eq:gamma_constraints} and \refeq{eq:DL_condition}, are correct, then we can generate statistically valid synthetic data, i.e., stationary and homogeneous data, while keeping the computation resource minimal. Here, we examine the conditions on $\DL$ with synthetic data using the auto-correlation function because this is the easiest way to see the effect of the finite spatial domain, and whilst the correlation lengths and times will also be affected these require fitting functions that add unnecessary complexity to the problem. We do not examine the $\DT$ condition because synthetic data are usually generated with many time points such that ensemble average can be performed, in which case \refeq{eq:gamma_constraints} is readily satisfied.

We generate synthetic data  with $\lx = 0.1$ m, $\tlife=15~\mu$s, $v=5,000$ m/s, $\DT=48,000~\mu$s and $\DTsub=480~\mu$s. We set the variance of amplitudes, $A^2$ in $P\lp A_i\rp$ (see \refeq{eq:xtA_dist}), constant in space with the intention of generating homogeneous synthetic data. To cover a couple of correlation lengths, we set $\DLv=0.2$ m. Based on \refeq{eq:DL_condition}, we find that $\DL\ge0.7$ m for this case. 

\reffig{fig:Fluctuation-level} shows, as functions of spatial position $x_a$, the $\lab C_{a, a}\rab^{1/2}$ (squares), i.e., the calculated fluctuation levels using \refeq{eq:corr_anal_final} with the actual values of $\WDL$ (no approximation of $\WDL$ to two) for $\DL=0.2$ (blue), $0.4$ (yellow), $0.8$ (green) and $1.4$ m (red). Note that the green squares are not visible as they are overlapped with the red squares. Numerically estimated fluctuation levels, $\lab\CaaSyn\rab^{1/2}$, based on four sets of synthetic data  with $\DL=0.2$ (blue dot), $0.4$ (yellow dash dot), $0.8$ (green dash) and $1.4$ m (red line) are also shown. We see that the data are not homogeneous for the cases of $\DL=0.2$ and $0.4$ m which do not satisfy the $\DL$ condition set by \refeq{eq:DL_condition} even if $A^2$ is set to be constant in space while generating the synthetic data. The underestimation of the fluctuation level towards the edge of $\DLv$ for these cases are caused by the `edge effect,' i.e., $\WDL<2$ towards the edge.  Data are homogeneous for $\DL=0.8$ and $1.4$ m, i.e., $\DL\ge0.7$ m is satisfied.

\begin{figure}[t]
\includegraphics[width=\linewidth]{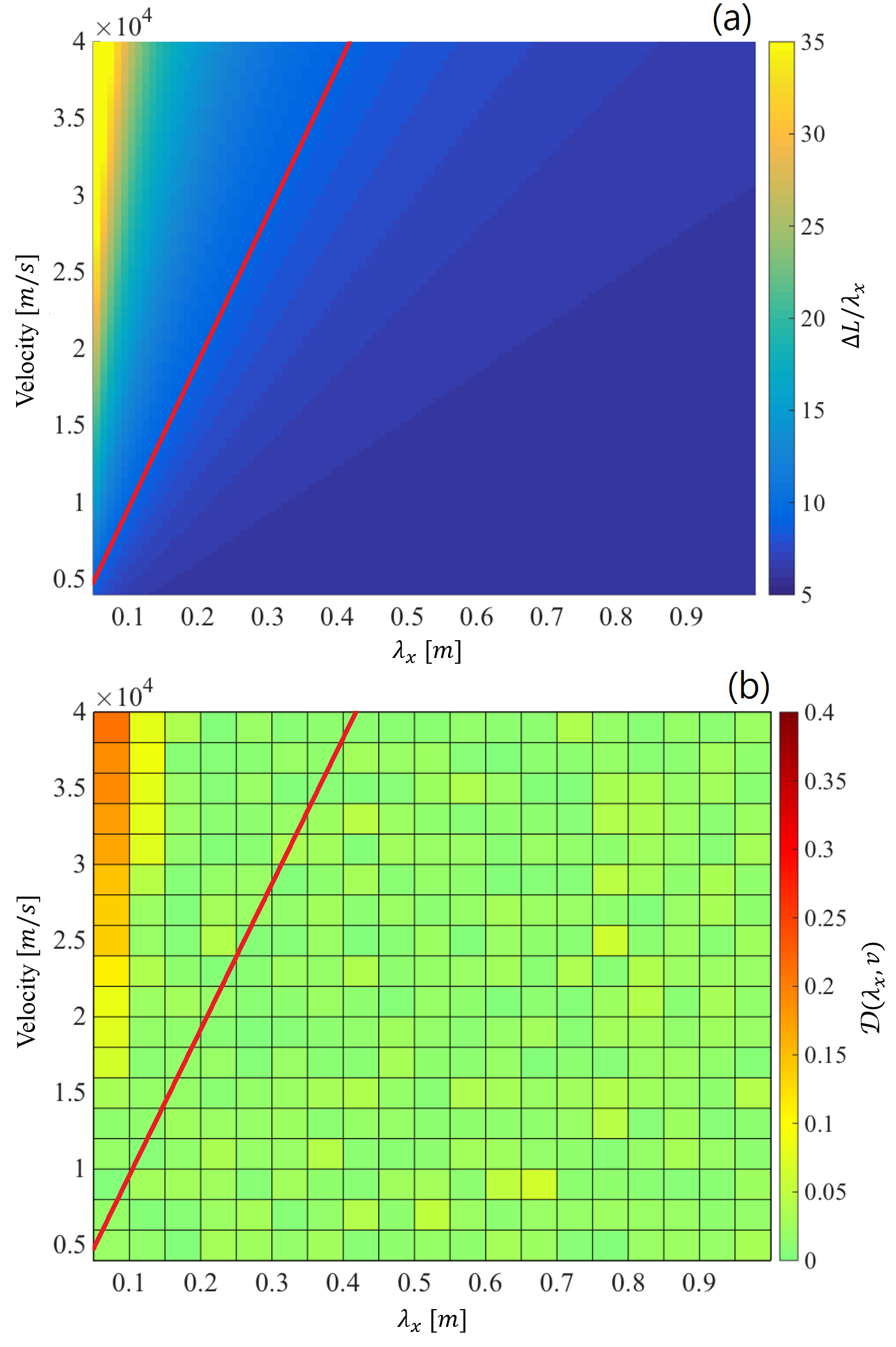}
\caption{(a) Required minimal $\DL$ normalized to $\lx$ as a function of $\lx$ and $v$ with fixed $\tlife=15~\mu$s and $\DLv=2\lx$. (b) The normalized average distance $\normdist$, defined in \refeq{eq:def_dist}, for $342$ ($18~\times~19$) sets of synthetic data with various values of $\lx$ and $v$ while keeping $\DL/\lx=9$ (red lines in both figures). If $\normdist$ is not close to zero, then the synthetic data may not be regarded as homogeneous, and the region violating the required $\DL$ condition, i.e., above the red line, has values of $\normdist$ greater than zero.}
\label{fig:Required}
\end{figure}

\reffig{fig:Required}(a) shows the required minimum $\DL$ normalized to $\lx$, calculated with \refeq{eq:DL_condition}, as a function of $\lx$ and $v$ while keeping $\tlife = 15~\mu$s with $\DLv=2\lx$. We, then, generate 342 sets of synthetic data with various values of $\lx$ and $v$ with the domain size of $\DL/\lx=9$ shown as the red line in \reffig{fig:Required}(a). Thus, we expect that the sets of synthetic data which fall into the region above the line of $\DL/\lx=9$ do not satisfy the homogeneous condition, i.e., these sets of data require a larger $\DL$. With the line (red) of $\DL/\lx=9$, \reffig{fig:Required}(b) shows the normalized average distance $\normdist$ of the numerically estimated fluctuation levels $\lab\CaaSyn\rab^{1/2}$ from the expected value $\lab C_{a, a}\rab^{1/2}$ (\refeq{eq:corr_anal_final}) defined as
\begin{equation}
\label{eq:def_dist}
\normdist=\dfrac{\sqrt{\dfrac{1}{N_{x_a}} \displaystyle\sum_{x_a} \lsb\lab\CaaSyn\rab^{1/2}-\lab C_{a, a}\rab^{1/2}\rsb^2}}{\lab C_{a, a}\rab^{1/2}},
\end{equation}
where the sum is performed on the all spatial positions within the $\DLv$, and $N_{x_a}$ is the number of spatial points. $\normdist$ is a zeroth order proxy for the homogeneity of synthetic data (see \reffig{fig:Fluctuation-level}). If $\normdist$ is not close to zero, then we speculate that the data are not homogeneous. \reffig{fig:Required}(b) clearly shows that $\normdist$ is conceivably larger than zero above the line, hence vindicating our proposed condition on $\DL$.

\section{Conclusion}
\label{sec:conclusion}

Motivated by the recent trend of wide usage of synthetic data either in a direct comparison of data from a local turbulence simulation to experimental data or in evaluating the reliability of a statistical algorithm, we have investigated how the synthetic data must be generated while minimizing the computation resource. The conditions on the total simulation domains, $\DT$ and $\DL$, given the `viewing' domain $\DLv$ (or the domain of interest) are summarized in \refeq{eq:gamma_constraints} and \refeq{eq:DL_condition} such that the data are stationary and homogeneous. We emphasize that many statistical analyses require the data to be homogeneous and stationary at least within the domain of interest. Furthermore, if one generates a synthetic data based on a local turbulence simulation such as a gyro-kinetic simulation, the conditions can be applied to $\DLv$ given $\DT$ and $\DL$.

We found the conditions by realizing that two error functions $\WDTsub$ in \refeq{eq:corr_aibi} and $\WDL$ in \refeq{eq:corr_ensemble}, which take roles of weighting factors on each eddy, must be approximately two throughout the domain of interest. As these error functions are manifestations of the coherent structure of Gaussian-shaped eddies, similar weighting factors would appear for different shapes of eddies as long as the eddies have non-zero coherent structures in temporal and/or spatial domains. Although we do not provide exact forms of weighting factors for different shapes of eddies as we have done for Gaussian-shaped eddies in this paper, our rationale can be applied to other shapes of eddies. We do not extend the conditions to include other shapes of eddies here because 1) the purpose of this paper is not to provide the conditions for all possible shapes of eddies (which is not possible), rather to provide the `rationale' how one must choose a domain of interest such that the synthetic data can be used for statistical analyses and 2) we have chosen to apply our rationale on the Gaussian-shaped eddies as we believe that such a shape is a good approximation \cite{turbulence} to describe turbulent fluctuations.

\section*{Acknowledgement}
This work is supported by National R\&D Program through the National Research Foundation of Korea (NRF) funded by the Ministry of Science, ICT \& Future Planning (grant number 2014M1A7A1A01029835) and the KUSTAR-KAIST Institute, KAIST, Korea.

\appendix
\section{Constant $\WDL$ within the `viewing' domain $\DLv$}
\label{app:homogeneous_state}

\begin{figure}[t]
\includegraphics[width=\linewidth]{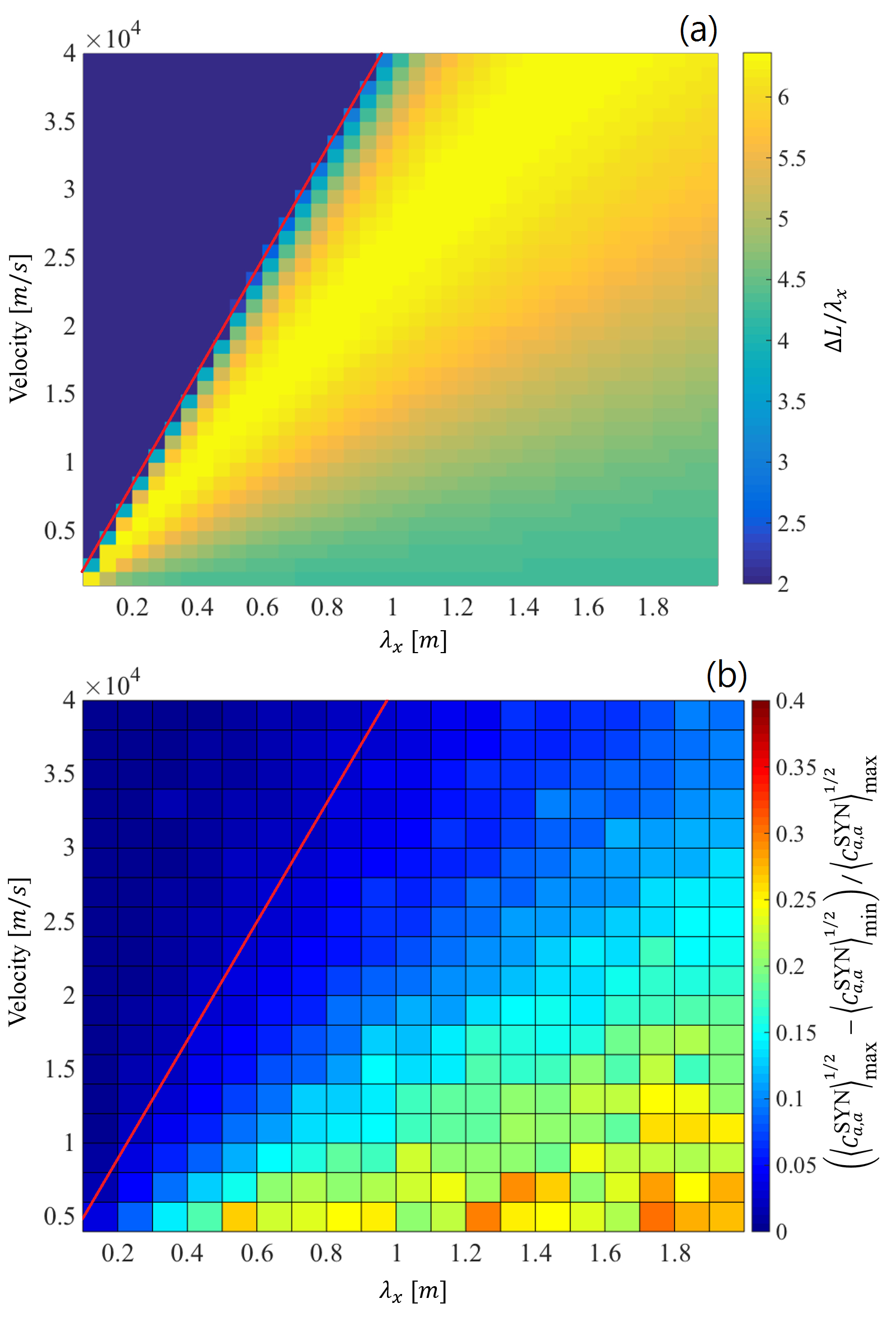}
\caption{(a) The minimum size of $\DL$ normalized to $\lx$ such that $\WDL$ is almost constant within $\DLv$, i.e., satisfying \refeq{eq:homo_like_cond}. Note that if this calculated $\DL$ is less than $\DLv$, then we force $\DL=\DLv$. (b) Normalized difference between the maximum and minimum fluctuation levels from the synthetic data with $\DL/\lx=2$ (red lines in both figures). The region above the line has constant $\WDL$ within $\DLv$, thus the fluctuation levels are almost constant.}
\label{fig:homo_like}
\end{figure}

$\WDL$ within the viewing window $\DLv$ can be stated as constant if the difference between the maximum and the minimum of $\WDL$ denoted as $\WDLmax$ and $\WDLmin$, respectively, are small (for instance, less than $5$\%): 
\begin{equation}
\label{eq:homo_like_cond}
\dfrac{\WDLmax-\WDLmin}{\WDLmax}\le 0.05.
\end{equation}
Here, we provide the definition of $\WDL$ again as a matter of convenience:
\begin{equation}
\label{eq:ensemble_corr_erf_app}
\WDL=\Erf\lsb\dfrac{x_{a}+x_{b}+\DL}{2\sqrt{\lx^{2}+\tlife^{2}v^{2}}}\rsb-\Erf\lsb\dfrac{x_{a}+x_{b}-\DL}{2\sqrt{\lx^{2}+\tlife^{2}v^{2}}}\rsb.
\end{equation}
To estimate the fluctuation level, we set $x_a=x_b$ as in $\lab C_{a, a}\rab^{1/2}$. By taking the first and the second derivatives of $\WDL$ with respect to $x_a$, we find that the maximum point is at $x_a=0$. Furthermore, because $x_a=0$ is the only critical point, the minimum occurs at the boundary of the $\DLv$. Then, we have
\begin{equation}
\label{eq:valus_WDL_max_min}
\begin{aligned}
\WDLmax =& \Erf\lsb\dfrac{\DL}{2\sqrt{\lx^{2}+\tlife^{2}v^{2}}}\rsb-\Erf\lsb\dfrac{-\DL}{2\sqrt{\lx^{2}+\tlife^{2}v^{2}}}\rsb, \\
\WDLmin =& \Erf\lsb\dfrac{\DLv+\DL}{2\sqrt{\lx^{2}+\tlife^{2}v^{2}}}\rsb-\Erf\lsb\dfrac{\DLv-\DL}{2\sqrt{\lx^{2}+\tlife^{2}v^{2}}}\rsb. \\
\end{aligned}
\end{equation}
Thus, the data may resemble the condition of homogeneous state if \refeq{eq:homo_like_cond} is satisfied with \refeq{eq:valus_WDL_max_min}.

\reffig{fig:homo_like}(a) shows, as a function of $\lx$ and $v$, the minimum size of $\DL$ normalized to $\lx$ satisfying \refeq{eq:homo_like_cond} with fixed $\tlife=100~\mu$s and $\DLv=2\lx$. Note that if this calculated minimum size of $\DL$ happens to be less than $\DLv$, we force $\DL=\DLv$ since it is non-sense to consider the larger `viewing' window than the total simulation window. Then, we generate 342 sets of synthetic data with $\DL/\lx=2$, i.e., $\DL=\DLv$. The lines of $\DL/\lx=2$ are depicted as red lines in both \reffig{fig:homo_like}(a) and (b). \reffig{fig:homo_like}(b) shows that in the region above the line of $\DL/\lx=2$, where \refeq{eq:homo_like_cond} is satisfied, the maximum and minimum fluctuation levels from the synthetic data, $\lab\CaaSyn\rab^{1/2}_\text{max}$ and $\lab \CaaSyn\rab^{1/2}_\text{min}$, respectively, within the $\DLv$ are quite similar; while the region below the line shows non-negligible differences.

\begin{figure}[t]
\includegraphics[width=\linewidth]{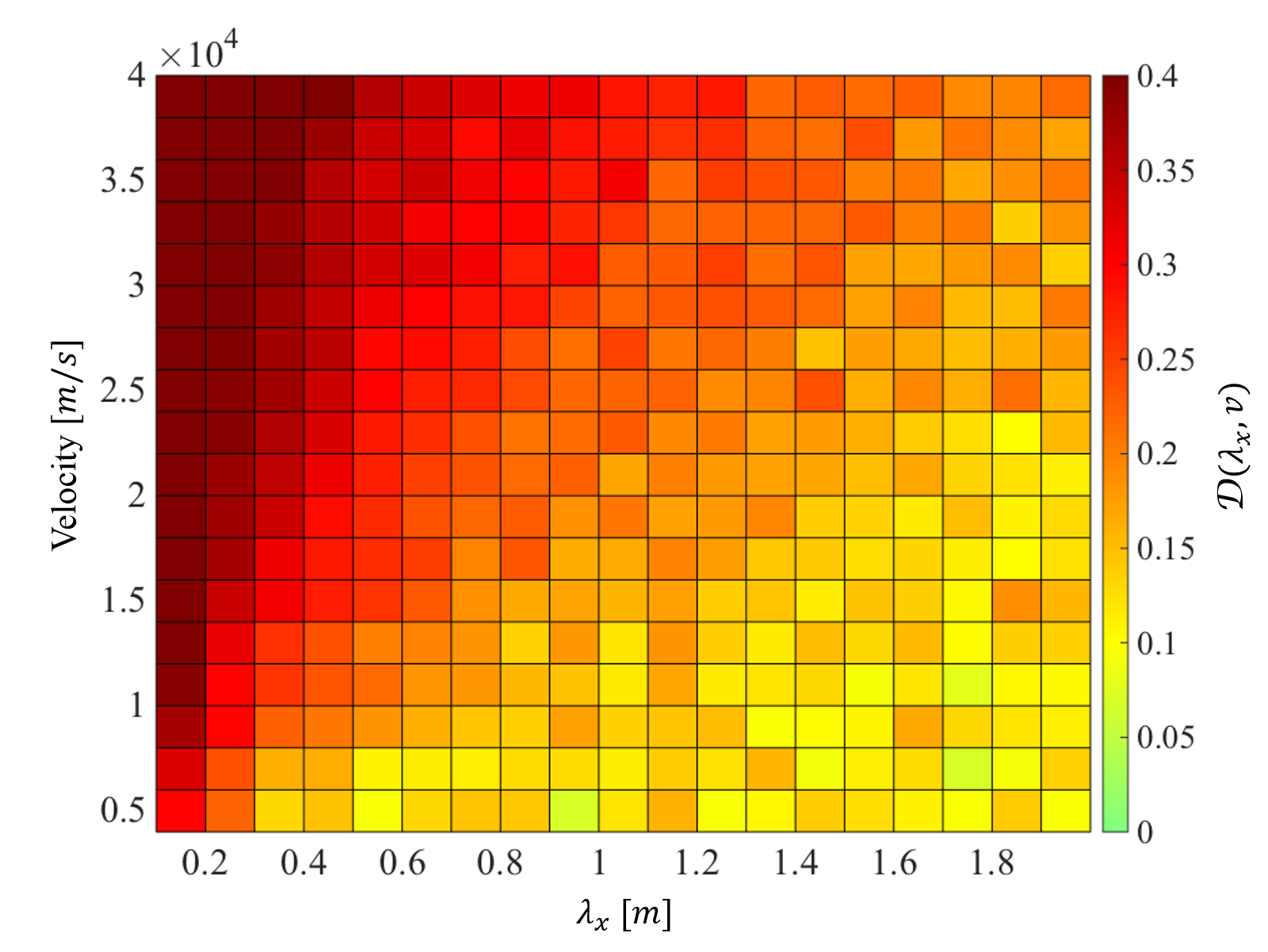}
\caption{Normalized average distance $\normdist$ defined in \refeq{eq:def_dist} for the 342 sets of synthetic data used to generate \reffig{fig:homo_like}. None of the data sets have the values of $\normdist$ close to zero as in \reffig{fig:Required}.}
\label{fig:difference_app}
\end{figure}

Although the fluctuation levels are constant over the viewing window for the data sets satisfying \refeq{eq:homo_like_cond}, we can easily find that none of the data sets in \reffig{fig:homo_like} satisfy the `true' homogeneous condition set by \refeq{eq:DL_condition}. If we plot the normalized average distance $\normdist$ defined in \refeq{eq:def_dist} for the same sets of the synthetic data, we see that all the synthetic data have values of $\normdist$ greater than zero as shown in \reffig{fig:difference_app} (cf. \reffig{fig:Required}(b) where we have deliberately used the same color scale).

As a summary, we state that constant $\WDL$ is not a sufficient condition for data being homogeneous, rather we require $\WDL\approx 2$.





\bibliographystyle{elsarticle-num}

\end{document}

%% file: my_macro.tex
\usepackage{color}

\newcommand{\lp}{\left(}
\newcommand{\rp}{\right)}
\newcommand{\lab}{\left<}
\newcommand{\rab}{\right>}
\newcommand{\lsb}{\left[}
\newcommand{\rsb}{\right]}

\newcommand{\labs}{\left|}
\newcommand{\rabs}{\right|}

\newcommand{\refeq}[1]{Eq. (\ref{#1})}
\newcommand{\refsec}[1]{Sec. \ref{#1}}

\newcommand{\reffig}[1]{Fig. \ref{#1}}

\newcommand{\abs}[1]{\labs #1 \rabs}

\newcommand{\KAISTNQe}{Department of Nuclear and Quantum Engineering, KAIST, Daejeon 34141, Korea}

\newcommand{\NFRI}{National Fusion Research Institute, Daejeon, Republic of Korea}

\newcommand{\culham}{EURATOM/CCFE Fusion Association, Culham Science Centre, Abingdon, OX14 3DB, UK}
\newcommand{\oxford}{Rudolf Peierls Centre for Theoretical Physics, University of Oxford, Oxford, OX1 3NP, UK}
\newcommand{\merton}{Merton College, Oxford, OX1 4JD, UK}


%% file: Jaewook_CPC2015.bbl
\begin{thebibliography}{99}


\bibitem{Carreras1997} B. Carreras, IEEE Trans. Plasma Sci. 25 (1997) 1281.
\bibitem{Burrell_ExB} K. H. Burrell, Phys. Plasmas 4 (1997) 1499.
\bibitem{Diamond_zonalflows2005} P. H. Diamond, S.-I. Itoh, K. Itoh, and T. S. Hahm, Plasma Phys. Controlled Fusion 47 (2005) R35.
\bibitem{Hahm_flowshear1995} T. S. Hahm and K. H. Burrell, Phys. Plasmas 2 (1995) 1648.
\bibitem{Huld_coherentstructure1991} T. Huld, A. H. Nielsen, H. L. Pecseli, and J. Juul Rasmussen, Phys. Fluids B 3 (1991) 1609.
\bibitem{Mckee_velocimetry2004} G. R. McKee, R. J. Fonck, D. K. Gupta, D. J. Schlossberg, M. W. Shafer, C. Holland, and G. Tynan, Rev. Sci. Instrum. 75 (2004) 3490.
\bibitem{ghim_prl_2013} Y.-c. Ghim, A. A. Schekochihin, A. R. Field, I. G. Abel, M. Barnes, G. Colyer, S. C. Cowley, F. I. Parra, D. Dunai, S. Zoletnik, and the MAST Team, Phys. Rev. Lett. 110 (2012) 145002.
\bibitem{ghim_nf_2014} Y.-c. Ghim, A. R. Field, A. A. Schekochihin, E. G. Highcock, C. Michael, and the MAST Team, Nucl. Fusion 54  (2014) 042003.
\bibitem{Levinson1984} S. Levinson, J. Beall, E. Powers, and R. Bengtson, Nucl. Fusion 24  (1984) 527.
\bibitem{Ghim2012} Y. c Ghim, A. R. Field, D. Dunai, S. Zoletnik, L. Bardo ́czi, A. A. Schekochihin, and the MAST Team, Plasma Phys. Controlled Fusion 54 (2012) 095012.
\bibitem{Durst1992} R. D. Durst, R. J. Fonck, G. Cosby, H. Evensen, and S. F. Paul, Rev. Sci. Instrum. 63 (1992) 4907.
\bibitem{Fonck1993} R. J. Fonck, G. Cosby, R. D. Durst, S. F. Paul, N. Bretz, S. Scott, E. Synakowski, and G. Taylor, Phys. Rev. Lett. 70 (1993) 3736.
\bibitem{White_pop_2008} A. E. White, L. Schmitz, G. R. McKee, C. Holland, W. A. Pee- bles, T. A. Carter, M. W. Shafer, M. E. Austin, K. H. Burrell, J. Candy, J. C. DeBoo, E. J. Doyle, M. A. Makowski, R. Prater, T. L. Rhodes, G. M. Staebler, G. R. Tynan, R. E. Waltz, and G. Wang, Phys. Plasmas 15 (2008) 056116. 
\bibitem{Shafer_PoP_2012} M. W. Shafer, R. J. Fonck, G. R. McKee, C. Holland, A. E. White, and D. J. Schlossberg, Phys. Plasmas 19 (2012) 032504.
\bibitem{Field_ppcf_2014} A. R. Field, D. Dunai, Y.-c. Ghim, P. Hill, B. McMillan, C. M. Roach, S. Saarelma, A. A. Schekochihin, S. Zoletnik, and the MAST Team, Plasma Phys. Controlled Fusion 56 (2014) 025012.
\bibitem{Bencze2005} A. Bencze and S. Zoletnik, Phys. Plasmas 12 (2005) 052323.
\bibitem{Balazs2011} B. Tal, A. Bencze, S. Zoletnik, G. Veres, and G. Por, Phys. Plasmas 18 (2011) 122304.
\bibitem{Guszejnov2013} D. Guszejnov, A. Bencze, S. Zoletnik, and A. Kra ̈mer-Flecken, Phys. Plasmas 20 (2013) 062303.
\bibitem{Mckee2003} G. R. McKee, C. Fenzi, R. J. Fonck, and M. Jakubowski, Rev. Sci. Instrum. 74 (2003) 2014. 
\bibitem{randomdata} J. S. Bendat and A. G. Piersol, Random Data - Analysis and Measurement Procedures (4th ed), Wiley, 2010.
\bibitem{06MIT_Carl} C. E. Rasmussen and C. K. I. Williams, Gaussian Processes for Machine Learning, MIT Press, 2006.
\bibitem{turbulence} P. Davidson, Turbulence - An introduction for scientists and engineers, Oxford, 2004.

\end{thebibliography}
